\documentclass[11pt,a4paper,superscriptaddress,prb,amsmath,amssymb,aps]{revtex4}
\usepackage{graphicx,graphicx,epsfig}

\usepackage{graphicx}
\usepackage{bm}

\begin{document}
\title{Reversible stochastic pump currents in interacting nanoscale conductors.} 

\author{N.A. Sinitsyn}
\affiliation{Center for Nonlinear Studies and Computer, Computational 
  and Statistical Sciences Division, Los Alamos National Laboratory,
  Los Alamos, NM 87545 USA}


\begin{abstract}
  I argue that the geometric phase, responsible for 
reversible pump currents in classical stochastic kinetics, can be observed experimentally with an electronic setup, similar to the ones
reported recently in {\em Phys. Rev. Lett.} {\bf 96}, 076605 (2006) and {\em Nature Physics} {\bf 3}, 243 (2007).  
\end{abstract}

\date{\today}

\maketitle 

The stochastic pump effect manifests itself during
periodic driving of a classical stochastic system, such as the enzyme
in the sea of interacting substrate and product molecules or ion channel in the
cell membrane  \cite{westerhoff-86}.
Usually the driving is achieved by an application of a time-dependent periodic electric field
that modulates chemical potentials and kinetic rates. As a result of time-dependent driving, part of the flux 
appears to have properties that have
no analog under the purely stationary conditions. In the adiabatic limit
this extra contribution to the flux  is reversible, i.e. it changes sign under the time reversal of the external
perturbation. 
 
Recently, it was discussed that the purely classical 
adiabatic pump effect has geometric origins
\cite{sinitsyn-07}, namely it is related to the geometric phase gained by
the flux moments generating function (mgf) under an external periodic driving
of kinetic rates. 
The theory in \cite{sinitsyn-07} allows to calculate both average geometric fluxes and their fluctuations.
While the experimental demonstration of the classical stochastic pump effect has been reported e.g. in
ion channel experiments \cite{tsong}, the specific geometric properties such as the Berry
curvature in the parameter space or the effect of the phase mismatch of driven kinetic rates
have not been measured. 
 
So far only the average pump fluxes have been studied experimentally. The geometric phase \cite{sinitsyn-07}, however,
contains much more information, so that its detailed experimental evaluation requires the derivation of the 
full counting statistics of pump currents. This is a complicated task when measuring the fluxes through the setup discussed in \cite{sinitsyn-07}
or in the analogous electronic setup, with an electron transport only through a quantum dot in the Coulumb blockade regime because the corresponding currents are very weak.

In this note I propose that the full counting statistics of reversible stochastic currents
 can be
studied in another experimental setup, reported recently in \cite{gustavsson-06,sukhorukov-07}. 
The setup consists of a quantum dot coupled to a quantum point contact (QPC) in the ballistic transport regime.
The voltage applied to the QPC generates the current $J$. This current, however, is controlled by the charge inside the 
quantum dot, which changes the tunneling barrier at QPC due to the Coulomb potential. In the simplest realization, the quantum dot can only have either one or no electrons inside. The switching between those states, in turn, is
influenced by two gate voltages \cite{note-2}. Although experiments were performed at low temperatures, a sufficiently strong decoherence was assumed so that
the behavior of the charge in the dot was described by purely classical Markov dynamics with rates $\Gamma_1$ and $\Gamma_2$ of transitions 
respectively in or out of the dot.

Under the stationary conditions, in experiments \cite{gustavsson-06,sukhorukov-07}, the full counting
 statistics
of electrons transferred through the quantum point contact was measured.
The counting statistics of the current through the QPC is much easier to measure experimentally but it is different from the one due to the direct current through the quantum dot.
Thus one cannot directly apply the expressions for pump currents, derived in \cite{sinitsyn-07} to the currents through the QPC but rather should derive the geometric phase of the QPC-current separately, using the same 
approaches.

 This work contains the derivation of the full counting statistics of the reversible pump current through the QPC and provides expressions for a direct comparison with experimental data.
In addition I provide a simplified intuitive explanation of the effect, which cannot be found in previous publications.
  
Consider  the time-dependent  transition rates $\Gamma_{1}$, $\Gamma_2$, which can be
induced by a slow periodic modulation of gate voltages. We will assume the adiabatic limit, so that the driving frequency $\omega$
is small i.e. $\omega \ll \Gamma_i$, ($i=1,2$). As in \cite{sukhorukov-07} we will assume fast decoherence so that the classical stochastic 
dynamics can be sufficient to describe the experiment.

The complete information about the flux through the QPC is contained in the moments generating function (mgf), defined by 
\begin{equation}
U(\lambda)=
\sum_{s=-\infty}^{\infty} P_{n=s}e^{is\lambda},
\label{pgf1}
\end{equation}
where $n$ is the number of electrons passed through the QPC (reverse transitions are counted with the negative sign).
Suppose that we know the mgfs of the currents under stationary conditions, when in addition the state of the dot is specified to be always either with
or without the electron inside during the whole measurement. For the time of measurement $T$ such mgfs can be written in the form
\begin{equation}
U_{i}(\lambda)=e^{TH_{i}(\lambda)},\,\,\,\,\,\, (i=1,2),
\label{pgf2}
\end{equation}
where $i=1,2$ correspond here respectively to the case without and with an electron inside the dot. 
Functions $H_i(\lambda)$ do not depend on $T$ if the latter is sufficiently large. Derivatives of $H_i(\lambda)$ provide cumulants of 
current distributions.

Now we allow transitions between empty and filled states of the dot with time-dependent rates $\Gamma_i(t)$. According to \cite{sukhorukov-04,sukhorukov-07} the
mgf satisfies the equation
 \begin{equation}
\partial_tU(\lambda)={\bf H} U,
\label{pgf3}
\end{equation} 
where the "Hamiltonian" ${\bf H}$ is given by \cite{note}
 \begin{equation}
{\bf H}(\lambda,t)=\left( \begin{array}{ll}
H_1(\lambda)-\Gamma_1(t)                  & \,\,\,\,\,\,\,\,\, \Gamma_2(t) \\
\,\,\,\,\,\,\,\, \Gamma_1(t)              & H_2(\lambda)-\Gamma_2(t) 
\end{array}\right),
\label{pgf4}
\end{equation} 
and unlike \cite{sukhorukov-07} we allow for the slow time dependence of parameters $\Gamma_i$. Also, unlike \cite{sinitsyn-07} the Hamiltonian (\ref{pgf4}) contains 
the counting parameter $\lambda$ at the main diagonal, rather than at off-diagonal matrix elements. This reflects the fact that the current through the QPC "counts" time of the
dot being in one of the states, rather than the current through the dot.

The evolution 
equation (\ref{pgf3}) is similar
to the evolution equation of spin-1/2 in time-dependent Zeeman field. Although the Hamiltonian (\ref{pgf4}) is not Hermitian, this analogy
can be employed to find the mgf $U(\lambda)$ in the adiabatic limit. Following the discussion in \cite{sinitsyn-07,nazarovFCS} the result can be expressed
as an exponent of the sum of the geometric and the quasistationary contributions

\begin{equation}
U(\lambda)=e^{S_{geom}(\lambda)+S_{qst}(\lambda)},
\label{pgf5}
\end{equation}
where $S_{geom}$ and $S_{qst}$ can be expressed in terms
of the instantaneous eigenvalue $h(\lambda,t)$ of the matrix (\ref{pgf4}) with the larger real part and the corresponding
 right and left instantaneous eigenvectors 
$|u(\lambda,t)\rangle $ and $\langle u(\lambda,t)|$. 

\begin{equation}
S_{qst}(\lambda)=\frac{T}{T_0} \int_0^{T_0}h(\lambda,t) dt ,
\label{pgf6}
\end{equation}
\begin{equation}
S_{geom}(\lambda)=-\frac{T}{T_0} \int_0^{T_0}\langle u(\lambda,t)|\partial_t |u(\lambda,t)\rangle dt,
\label{pgf7}
\end{equation}
where $T_0=2\pi/\omega$.
The quasi-stationary contribution (\ref{pgf6}) is merely the time-average of the stationary counting statistics, derived in \cite{sukhorukov-07}, while
the geometric part (\ref{pgf7}) is a new term, that has no analog in the steady state.
Next we will use the fact that the time-dependence of $| u(\lambda,t) \rangle$ is due to the time-dependence of parameters $\Gamma_i(t)$ only, which allows to 
rewrite the geometric contribution in terms of the circulation of the vector ${\bf A}$, $A_i=\langle u|\partial_{\Gamma_i} u\rangle$ along the contour ${\bf c}$
in the parameter space
or equivalently, as
the integral of the 2-form $F_{1,2}=\langle \partial_{\Gamma_1} u|\partial_{\Gamma_2} u\rangle-
\langle \partial_{\Gamma_2} u|\partial_{\Gamma_1} u \rangle$ over the surface ${\bf s_c}$ inside this contour. Substituting expressions for eigenvectors and the
eigenvalue of (\ref{pgf4}) into (\ref{pgf6}) and (\ref{pgf7}) we find

\begin{equation}
S_{{\rm geom}}(\lambda)=-\frac{T}{T_0}\oint_{{\bf c}}{\bf A} \cdot d {\bf \Gamma} =-\frac{T}{T_0} \int_{{\bf s_c}} d\Gamma_1 d\Gamma_{2} F_{1,2},
\label{sgeom}
\end{equation}
\begin{equation}
F_{1,2}= \frac{H_2(\lambda)-H_1(\lambda)}
{[K^2-4(H_1(\lambda)\Gamma_2+H_2(\lambda)\Gamma_1-
H_1(\lambda)H_2(\lambda))]^{3/2}},
\label{BC}
\end{equation}
\begin{equation}
\begin{array}{l}
S_{\rm qst}(\lambda)=\frac{T}{2T_0} \int_0^{T_0} dt \{ K+ \\
\\ 
  \sqrt{ K^2+ 
4[\Gamma_2H_1(\lambda)+
\Gamma_1H_2(\lambda)-
H_1(\lambda)H_2(\lambda)] }   \},
\end{array}
\label{qst}
\end{equation}
where we introduced the vector ${\bf \Gamma}=(\Gamma_1,\Gamma_2)$
and $K=K({\bf \Gamma},\lambda)=H_1(\lambda)+H_2(\lambda)-\Gamma_1-\Gamma_2$.
The 2-form $F_{1,2}=F_{1,2}({\bf \Gamma},\lambda)$ is an analog of the Berry
curvature in quantum mechanics. It is responsible for the 
reversible component of the current.

For a strong current through the QPC, as it is discussed in \cite{sukhorukov-07},
one can disregard the noise part of $H_i(\lambda)$, in comparison to the noise due to
interactions with the quantum dot i.e. one can use the simplified form $H_1=iI_1 \lambda$ and $H_2=iI_2 \lambda$, where
$I_1,I_2$ are currents through the QPC respectively when the dot is empty and filled with an electron.
Now cumulants of the flux through the QPC can be found
by differentiating (\ref{sgeom}) and (\ref{qst}) with respect to $\lambda$.
Thus we find 
that the average current through the QPC is $J=J_{geom}+J_{qst}$, where 

\begin{equation}
  J_{\rm geom}=\int\int_{{\bf  s_c}} d\Gamma_1d\Gamma_2 
  \frac{I_1-I_2}{T_0\left[ \Gamma_1+\Gamma_2 \right]^3},
\label{JJ}
\end{equation}
\begin{equation}
J_{\rm qst}=\int_0^{T_0} dt\, \frac{\Gamma_1(t) I_2+\Gamma_2(t)I_1}{T_0\left[\Gamma_1(t)+\Gamma_2(t)\right]}.
\label{JJJ}
\end{equation}

The expressions for the average currents (\ref{JJ}) and (\ref{JJJ}) can be derived in a much more simplified way which, however, is not easy to apply to find higher cumulants. 
Let $P_e$ and $P_f=1-P_e$ are respectively probabilities of the dot to have no and have one electron inside.  $P_e$ satisfies a first order differential equation with the solution

\begin{equation}
 P_e(t)=\int^t_{t_0}dt' \Gamma_2(t')e^{-\int_{t'}^t (\Gamma_1(t'')+\Gamma_2(t''))dt''}.
\label{pe}
\end{equation}

Due to the fast decaying exponent, the integral (\ref{pe}) is dominated by the direct vicinity of the time point $t$ where we can approximate 
$\Gamma_i(t') \approx \Gamma_i(t) - (t-t')\partial_t \Gamma_{i}(t)$. Then integrating over time we find 
\begin{equation}
 P_e(t)\approx \frac{\Gamma_2(t)}{\Gamma_1(t)+\Gamma_2(t)} +{\bf a}\cdot{\bf \dot{\Gamma}}, 
\label{pe2}
\end{equation}
where ${\bf a}$ is the vector over the parameter space with components ${\bf a}=(\Gamma_2/(\Gamma_1+\Gamma_2)^3, -\Gamma_1/(\Gamma_1+\Gamma_2)^3)$.
Note that the vector field ${\bf a}={\bf a}({\bf \Gamma})$ has a nonzero vorticity
$\partial a_{\Gamma_2}/\partial \Gamma_1-\partial a_{\Gamma_1}/\partial \Gamma_2=1/(\Gamma_1+\Gamma_2)^3$, thus a circulation of ${\bf a}$ over a closed contour can be nonzero.
 The average current is
\begin{equation}
 J=I_1P_e+I_2P_f.
\label{avcur}
\end{equation}
Substituting (\ref{pe2}) into (\ref{avcur}) and averaging over the period of the parameter modulation, one will recover (\ref{JJ})  and (\ref{JJJ}).

Can the reversible current be observed 
experimentally? 
Generally the geometric contribution is much weaker than the quasistationary one. In the adiabatic limit it is suppressed by the ratio $\omega/\Gamma_i \ll 1$.
However, the specific symmetry of this contribution provides the opportunity to detect it. The geometric part of the full counting statistics $S_{geom}$ changes sign
under the change of the direction of "motion" along the contour ${\bf c}$, while the quasistationary part remains the same. This suggests the obvious 
strategy  to extract $S_{geom}$ experimentally, namely one should perform the measurements of the mgf under periodically time-dependent gate voltages and then
to perform the same type of measurements during the same period of time but for the time-reversed perturbation. 

For example, if during the first experiment one
drives the rates according to the law $\Gamma_1(t)=a+b$cos$(\omega t)$ and $\Gamma_2(t)=c+d$cos$(\omega t+\phi)$ with constants
 $a,b,c,d$ and the phase mismatch $\phi$, then the second measurement should be done for the driving 
with the opposite sing of the phase mismatch, namely such that $\Gamma_1(t)=a+b$cos$(\omega t)$ and $\Gamma_2(t)=c+d$cos$(\omega t-\phi)$. Taking the 
difference between two corresponding counting statistics, the quasistationary contributions cancel but the geometric contributions, being different only by the sign, do not
cancel and make the final result of such a measurement equal to $2S_{geom}(\lambda)$. We note that to observe it the driving of the rates should be out of phase.
This is clear from the fact that the geometric contribution is finite when the area inside the contour ${\bf c}$ is also finite, which can be achieved only when there is a phase
mismatch between $\Gamma_1$ and $\Gamma_2$ modulations. Changing this constant phase difference one can manipulate the strength of the reversible 
current contribution, for example, change its sign.  Fig 1. shows an example of the contour in the parameter space, leading to a nonzero pump current for $\phi=\pi/2$ and its time reversed
counterpart, corresponding to $\phi=-\pi/2$. The zero value of the reversible pump current can be achieved at $\phi=0$ or $\phi=\pi$.


\begin{figure}
\includegraphics[width=8.5 cm]{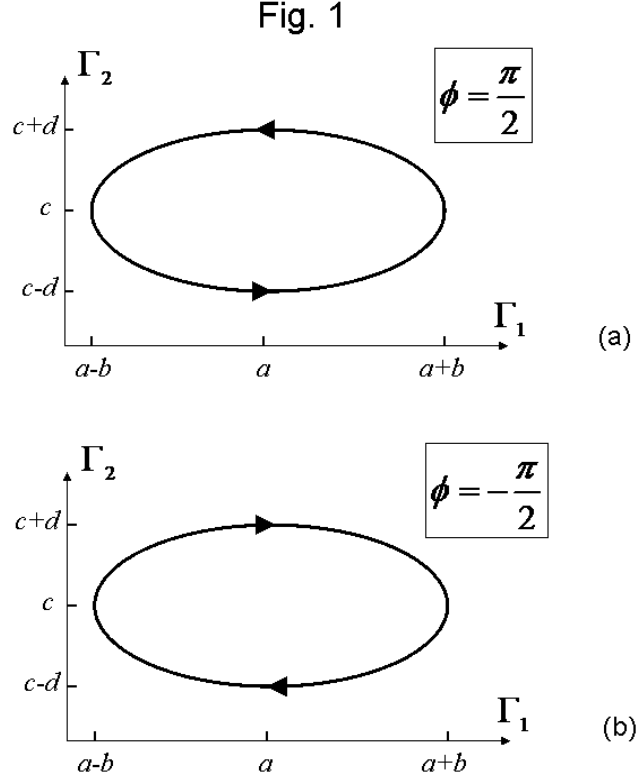}
\centering
\caption{\label{contour_dot} (a) A contour in the parameter space and (b) its time reversed counterpart, leading to nonzero reversible pump currents through the QPC.}
\end{figure}


The geometric phase can be clearly observable if the difference of
 the measured total transported charge during the forward and the time reversed modulations of gate voltages is much larger than the size of its typical fluctuations. 
The latter is dominated by a quasistationary shot noise, which can be estimated from the 2nd cumulant of the current 
$J^{(2)}\sim 2(I_1-I_2)^2\Gamma_1\Gamma_2/(\Gamma_1+\Gamma_2)^3$. Taking the data from experiment \cite{sukhorukov-07}: $\Gamma_i \sim 500$Hz,
time of measurement $T\sim 10^3$s, then assuming that the amplitude of modulation $\Delta \Gamma_i \sim 300$Hz and the modulation frequency
$1/T_0 \sim 50$Hz, we find the order of the signal/noise ratio $\eta=J_{geom}T/\sqrt{J^{(2)}T} \sim 10$, which 
can be good enough to confirm the presence of the effect. Note also 
that this ratio can be enhanced by increasing the measurement time ($\eta \sim T^{1/2}$).    

In conclusion, I examined the possibility to measure the geometric phase in the setup discussed in the recent work \cite{sukhorukov-07}. The
 estimates show that at least the average of the reversible current can be detected for a realistic choice of parameters. 
Such measurements of the Berry curvature  are important to enhance the control over the microscopic device with time-dependent perturbations.

\begin{acknowledgments} 
{\it  This work was funded in part by DOE under Contract No.\
  DE-AC52-06NA25396.} 
\end{acknowledgments}


\begin{thebibliography}{99}
\bibitem{note-2} Currents through the QPC in the system considered in \cite{gustavsson-06,sukhorukov-07} are similar, from the 
mathematical point of view, to the fluxes of a protein generated by a gene that stochastically flips between active and inactive states.
The role of gate voltages then is played by transcription factors, that regulate the gene's activity. Thus our mathematical
results also can be used in this biological context.
\end{thebibliography}
\end{document}